# All-Optical Arithmetic and Combinatorial Logic Circuits with High-Q Bacteriorhodopsin Coated Microcavities

**Sukhdev Roy[*], Mohit Prasad**
*Department of Physics and Computer Science*
*Dayalbagh Educational Institute (Deemed University)*
*Dayalbagh, Agra 282 110 INDIA*
*Ph.: +91-562-2801545, Fax: +91-562-2801226*
*[*]E -mail: sukhdevroy@dei.ac.in*

**Juraj Topolancik[1] and Frank Vollmer**
*Biofunctional Photonics Group*
*The Rowland Institute, Harvard University*
*100 Edwin H. Land Blvd., Cambridge, MA 02142U.S.A.*
*Ph.: +1-617 497 4681, Fax: +1-617-497-4627*
*E-mail: vollmer@rowland.harvard.edu*

*[1]Present Address: Department of Electrical and Computer Engineering*
*Northeastern University*
*360 Huntington Avenue, Boston, MA 02115*
*Ph.: +1-617-373-4159, Fax: +1-617-373-8970*
*E-mail: jtopolan@ece.neu.edu*

## Abstract

We present designs of all-optical computing circuits, namely, half-full adder/subtractor, de-multiplexer, multiplexer, and an arithmetic unit, based on bacteriorhodopsin (BR) protein coated microcavity switch in a tree architecture. The basic all-optical switch consists of an input infrared (IR) laser beam at 1310 nm in a single mode fiber (SMF-28) switched by a control pulsed laser beam at 532 nm, which triggers the change in the resonance condition on a silica beed coated with BR between two tapered fibers. We show that fast switching ~50 μs can be achieved by injecting a blue laser beam at 410 nm that helps in truncating the BR photocycle at the M intermediate state. Realization of all-optical switch with BR coated microcavity switch has been done experimentally. Based on this basic switch configuration, designs of all-optical higher computing circuits have been presented. The design requires $2^n-1$ switches to realize n bit computation. The proposed designs require less number of switches than terahertz optical asymmetric demultiplexer based interferometer designs. The combined advantages of high Q-factor, tunability, compactness and low power control signals, with the flexibility of cascading switches to form circuits, makes the designs promising for practical applications. The design combines the exceptional sensitivities of BR and microcavities for realizing low power circuits and networks. The designs are general and can be implemented (i) in both fiber-optic and integrated optic formats, (ii) with any other coated photosensitive material, or (iii) an externally controlled microresonator switch.

# I. INTRODUCTION

The anticipated requirement for ultrahigh speed ultrahigh bandwidth information processing has provided tremendous impetus for realization of all-optical devices and circuits. The advent of nano and bio technologies in the design, synthesis and characterization of novel materials and structures that include nanophotonics, biophotonics, plasmonics, organic and silicon photonics, photonic crystals and metamaterials, high-Q microresonators, slow and fast light and quantum information processing has opened exciting new possibilities for generation, manipulation and detection of light along with integration of multiple components and devices to achieve all-optical information processing [1-3].

Recent years have witnessed a renewed interest in optical computing due to two important advancements, namely, the development of (i) efficient conservative and reversible logic, algorithms and architectures that exploit the strengths of optics, and (ii) novel materials that exhibit high nonlinearities and structures that confine light to low dimensions that facilitates low power nonlinear optics [4-8]. These advancements offer tremendous scope to overcome the impediments to realize optical computing such as cascading components and devices to fabricate circuits and networks in the form of 2D/3D arrays that are small enough to have low switching energies and high speeds [9, 10].

The flexibility of electronic processing stems from its ability to perform non-linear operations such as thresholding. In optical processing, non-linear optical mechanisms play important role in ultrafast optical switches and all-optical logic gates. The integration of photonic components is expected to increase significantly due to emerging novel photonic structures such as microresonators, photonic crystals and plasmonics [11].

Microcavities have emerged as extremely sensitive and versatile device configurations for a variety of operations due to their high Q-factor, low switching threshold and ultra compactness [12, 13]. Optical microcavities confine light to small volumes by resonant circulation. For a coupled, input power $P_{in}$, the circulating intensity within the resonator is given by $I = P_{in} (\lambda/2\pi n) (Q/V)$ where n is the group index. For a cavity of Q $\sim 10^8$ and a mode volume of 500 $\mu m^3$ (both obtainable in spheres roughly 40 $\mu m$ in diameter), the circulating intensity exceeds 1 GW/cm$^2$ with less than 1 mW of coupled input power [12].

A very wide range of microresonator shapes have been explored over the years for various applications. The most widely used are rotationally symmetric structures such as Fabry-Perot cavities, spheres, cylinders, disks, torroids, and photonic crystals which have been shown to support very high-Q whispering gallery (WG) modes, whose modal field intensity distribution is concentrated near the dielectric-air interface [12-18].

Devices based on microcavities are already indispensable for a wide range of applications and studies. By tailoring the microcavity shape, size or material composition, the microcavity can be tuned to support a spectrum of optical modes with required polarization, frequency and emission patterns [12-15]. Some key optical microresonator material systems are Si, SOI, $SiO_xN_y$, GaAs, InP, GaN and $LiNbO_3$ [14]. Various all-optical logic operations have been shown using Si, GaAs and InGaAs microcavities [19-20]. The nonlinear optical mechanism implemented in the gates is the change in the refractive index from free charge carriers generated by two photon absorption (TPA). Although the change in refractive index in these microcavities is high, the absorption generates heat inside the microcavity due to pump power. The pump and probe power is in mWs and the Q-factor $\sim 10^4$ [19-20]. Recently, a spatially non-blocking optical router with a footprint of 0.07 mm$^2$ has been demonstrated [21]. The device is dynamically switched using thermo-optically tuned silicon microring resonators with a wavelength shift to power ratio of 0.5nm/mW. The design can route four optical inputs to four outputs with individual bandwidths of upto 38.5 GHz. These configurations can route a single-wavelength laser and provides a maximum extinction ratio larger than 20dB [21].

Silica microcavities have an inherent advantage of high Q-factor, relatively simple fabrication, possible on chip integration and control of the coupling efficiency through taper by the change of the fiber thickness [12-14, 17]. Fiber optic tapers have been proposed as a means to couple quantum states to or from a resonator onto a fiber [12]. Also the recent demonstration of a fiber-taper-coupled ultrahigh-Q microtoroid on a chip enables integration of wafer based functions with ultralow-loss fibre-coupled quantum devices. The bulk optical loss from silica is also exceptionally low and record Q factors of 8 x $10^9$ (and finesse of 2.3 * $10^6$) have been reported [12]. Ultra high-Q microtoroidal silica resonators represent a distinct class of optical microresonators with Q's in excess of $10^8$ [17]. Due to the high Q-factor and the small dimensions, switching at low power is feasible. Moreover, coating the microcavity with a photosensitive material can further lead to switching at ultra-low powers.

Recently, all-optical switching in the near infrared with bacteriorhodopsin (BR) coated silica microcavities has been reported, with a Q-factor on BR adsorption ~5 x $10^5$ [22]. The state of the BR in optical microcavities is controlled by a low power (< 200 μW) continuous green pump laser coupled to the microsphere cavity using a tapered fiber [22-24]. The photochromic protein bacteriorhodopsin (bR) found in the purple membrane fragments of *halobacterium halobium,* has emerged as an outstanding photonic material for practical applications due to its unique multifunctional photoresponse and properties [22-28]. By absorbing green–yellow light, the wild-type BR molecule undergoes several structural transformations in a complex photocycle that generates a number of intermediate states [29-32]. The main photocycle of BR is as shown in Fig.1 After excitation with green–yellow light at 570 nm, the molecules in the initial *B*-state get transformed into *J*-state within about 0.5 ps. The species in the *J*-state thermally transforms in 3 ps into the intermediate *K*-state, which in turn transforms in about 2 μs into the *L*-state. From the *L*-state, BR thermally relaxes to the *M*I-state within 8 μs and undergoes irreversible transition to the *M*II-state. The molecules then relax through the *N* and *O* intermediates to the initial *B*-state within about 10 ms. An important feature of all the intermediate states is their ability to be photo-chemically switched back to the initial *B*-state by shining light at a wavelength that corresponds to the absorption peak of the intermediate in question [29-32]. The wavelength in nm of the absorption peak of each species is shown as a subscript in Fig.1 Photoinduced molecular transitions in BR can be used to reversibly configure an ultrasensitive micron scale photonic component in which the optical response is resonantly enhanced.

Cascading to integrate a large number of all-optical components such as switches, logic gates etc., is a complex problem and a major obstacle in the development of a complete all-optical computing system [1, 8]. Branching of signals all-optically among various logic devices is a critical task. Theoretical schemes suggesting alternative solutions for parallel generation of logic gates have been recently reported in literature [33-35]. An effective method for cascading involves the tree architecture [36]. It is a multiplying system of a single straight path into several distributed branches and sub-branch paths. Shen and Wu have demonstrated that re-configurability can be introduced into designs of all-optical logic circuits by electo-optic switches [37]. Recently, Kodi and Louri have demonstrated that optical based system architecture shows better performance than electrical interconnects for uniform and non-uniform patterns without the application of reconfiguration techniques [38]. Even with the application of reconfiguration techniques, the dynamically reconfigurable optoelectronic provides much better performance for all communication patterns. Caulfield et al. have also proposed an electro-optical logic system with silicon-on-insulator (SOI) resonant structures [39]. Thus, low power BR based microcavity switch in a suitable architecture can be useful to achieve all-optical computing.

In this paper, we show fast switching (~50μs) in BR coated microcavity switch by injecting an blue laser beam (405 nm) in synchronization with green laser beam (532 nm) and further combine the advantages of BR coated microcavity switches in a tree architecture to realize reconfigurable all-optical logic and arithmetic operations. The key object of this paper is to propose general designs for optically controlled micrcrcavity based computing circuits, i.e., all-optical arithmetic and combinatorial circuits that can be configured to realize different logic and arithmetic operations in parallel. The whole logic unit consists of only BR coated microcavity switches and the horizontal and vertical extension of the designs can also be easily realized.

### II. EXPERIMENTAL SETUP OF BR COATED MICROCAVITY SWITCH

The basic configuration of the all-optical switch in the design has been considered to be the BR based microcavity switch [22]. The silica micro-cavity in contact between two tapered fibers serves as a four-port tunable resonant coupler. The diameter of the silica microcavity is 300 µm. The switch shown in Fig.2 represents the simplified schematic representation of the resonant coupler. Here, Port 1 acts as port for incoming as well as for control signal. Port 2 acts as lower output port and port 3 acts as an output port. The various states have been listed in Table I. In the BR coated microcavity switch, the switching of incoming signal operating at wavelength 1310 nm between the output ports (2 and 3) is photo-induced with a fiber-coupled green pump laser (at 532 nm) which controls the conformational state of the adsorbed BR. The molecularly functionalized microcavity thus redirects the flow of near-infrared light beam between two optical fibers. With the pump OFF, the probing light from input port 1 is detuned from resonance and is directly transmitted into the output port 2. The pump evanescently excites whispering gallery modes (WGM) propagating around the microsphere's equator, inducing photoisomerization along their path. A low green cw laser (< 200 µW at 532 nm) is sufficient for this purpose as its effective absorption is resonantly enhanced. Isomerization reduces the retinal polarizability, tuning the peak/trough of the resonance to match the wavelength of the infrared probe which is then rerouted into the output port 3 [22-24].

The schematic diagram of the experimental setup for all-optical switching with BR coated microcavity is shown in the Fig. 3. Resonant modes were excited with a distributed feedback laser, operating around 1310 nm, connected to port 1 via. single mode fiber (SMF-28, Dow-Corning, Midland MI) through fiber coupler. Green pump beam at wavelength 532 nm and a blue pump beam at wavelength 405 nm were also injected into port 1 of the switch. A 300 μm silica microsphere was formed by melting the tip of a single mode fiber in a butane nitrous oxide flame. Three layers of BR mutant D96N (Munich Innovative Biomaterials, Munich, Germany) were adsorbed onto the microsphere surface using alternate electrostatic deposition of cationic poly(dimethyldaiallyl)ammoniumchloride (PDAC) and anionic BR membranes. In each cycle a single oriented PDAC/BR monolayer was self-assembled onto the microcavity surface. The BR adsorption process slightly degraded the microcavity Q to ~5 x $10^5$, which was caused by the introduction of the scattering impurities during each drying process. Two parallel, single mode fibers held at 250 μm apart in a standard 1 cm acid resistant polystyrene cuvette were tapered by hydrofluoric acid erosion. Once etched, the fibers were immersed in 0.01 M phosphate buffered saline with pH = 7.4. The BR coated microsphere was then spring loaded between the two tapered fibers. To determine the resonant wavelength, the modulation current scanned periodically scanned at 100 Hz with a sawtooth shaped function. Photodiodes PD1, PD2 connected to fiber ports 2 and 3 were used to monitor the transmitted intensity of the probe and a spectrum containing 1000 points was recorded every ~200 ms with a LABVIEW program, as shown in the schematic diagram in Fig.3. The transmitted spectra exhibited an extinction of –9.4 dB in port 2 and a 9.8 dB increase in transmission in port 3. In order to achieve fast switching, a blue light beam is injected into the BR coated microcavity switch in synchronization with a green pump laser beam. The blue light beam operating at wavelength 405 nm helps in truncating the photocycle of the BR molecules at the M intermediate state, which is near to its peak absorption wavelength of 410 nm. Fig.4 shows the variation of the signal probe beam operating at 1310 nm in BR coated microcavity with the modulating signals green and blue pump beams operating at wavelengths 532 nm and 405 nm respectively. Switching takes place in μs as compared to the previous experimental results where switching occurs in ms.

### III. DESIGN OF ALL-OPTICAL COMPUTING CIRCUITS

### 1. HALF-FULL ADDER/SUBTRACTOR

A half-adder/subtractor is the basic building block of computing circuits. Fig.5 shows the architecture of a half-adder/subtractor based on BR coated microcavity switches S1, S2, and S3. Here X and Y act as optical control signals operating at wavelength 532 nm with a power of 200 μW. A laser source (LS) is coupled at port (1) of switch S1. Considering the detected output signal power ~ 200 μW, with an extinction of 9.8 dB [22], LS of 0.25 mW is enough to act as an incoming signal for the half-adder/subtractor. The cw light from the LS acts as an incoming signal to the switch S1. When the LS is switched OFF there is no light detected in any of the output ports of the architecture. When the LS is switched ON and both the optical control signals are zero i.e. X = 0 and Y = 0, no light is incident on the BR microcavity switches. BR is in its ground state hence, the incoming signal passes from the port (1) of switch S1 to port (1) of switch S3. As Y = 0, therefore light emerges from the port (2) of the switch S3 i.e. output port OP1. The logic generated at output port OP1= X′Y′, as shown in Table II.

Now, when X = 0 and Y = 1, light is incident on BR microcavity switches S2 and S3 respectively. It triggers the conformational changes in the BR. Thus, when an incoming signal is incident on port (1) of switch S1 it emerges from port (2) of switch S1, and reaches incoming port (1) of switch S3. Light at 532 nm is incident on switch S3; it redirects the incoming signal from port (1) to the port (3) of switch S3 via whispering gallery modes. The signal emerges from the output port OP2. The logic generated at output port OP2= X′Y, as shown in Table II.

Considering the values taken by optical control signals as X = 1, Y = 0, when cw light is incident on the incoming port of switch S1, presence of the light (at 532 nm) at switch S1 triggers the switch to route the incoming signal from the port (1) to the port (3) of switch S1. The incoming signal reaches port (1) of the switch S2. As Y = 0, no light is incident on the switch and hence the incoming signal is directed to the output port OP3, via port (2) of switch S2. Thus light emerges from the output port OP3. The logic generated at output port OP3= XY′, as shown in Table II.

Similarly, when optical control signals X = 1, and Y = 1 i.e. light (at 532 nm) is incident on all the switches. BR in all the microcavity switches will undergo conformational change and hence incoming signal is directed from port (1) to the port (3) of the switch S1. From there it reaches the incoming port (1) of switch S2. As Y = 1, the incoming signal is again rerouted to the port (3) of the switch S2 and hence to the output port OP4. The logic generated at output port OP4= XY, as shown in Table II.

Logic operations generated at output ports OP1, OP2, OP3 and OP4 are X′Y′, X′Y, XY′ and XY respectively as shown in Table II. Different logic operations can be derived by combining output of these ports. For e.g. if we combine signals of output ports OP1 with OP2 it results into logic operation X′, combining OP2 with OP3 results into logic operation of XOR. Sixteen different logic operations can be derived using this procedure. The logic operations are X, Y, X′, Y′ XOR′, XOR, X+Y′, X′+Y, X′+Y′, X+Y, X′Y, XY, X′Y′, XY′, T and F.

Working on this same principle, this tree architecture in combination with beam splitters and combiners can result in an all-optical half adder and subtractor as shown in Fig.5. Table II clearly depicts the various logic operations generated at output ports OP1, OP2, OP3, and OP4 respectively. Considering Table II carefully, if we combine light signals from output ports OP2 and OP3 through a beam combiner, it results into an XOR operation which is "sum" bit of an half adder. "Carry" bit is given by output port OP4. In the case of half subtractor, the "difference" bit is obtained by combining port OP2 and OP3, whereas "borrow" bit is given by the output port OP2. Truth table of half adder and subtractor is shown in Table III.

To implement tree architecture for triple-input-binary logic we have to incorporate another four BR based microcavity optical switches S4, S5, S6 and S7 as shown in Fig 6. There are eight different input combinations for implementing this logic. Depending on the state of input variables f (X, Y, Z) the output is obtained from output port OP1 to output port OP8. The truth table of full adder and subtractor for three input binary variables is given in Table IV. In the case of the full addition we have two outputs one is "sum" and another is "carry". Here "sum" takes the expression ∑f (1, 2, 4, 7) and "carry" takes the expression ∑f (3, 5, 6, 7). In the case of full subtraction we have two outputs, one is difference and the other is "borrow". Here difference can be expressed as ∑f (1, 2, 4, 7) and "borrow" can be written as ∑f (1, 2, 3, 7).

Tree architecture, which has three optical control signals X, Y and Z and eight outputs shown in Fig. 6, can successfully be used to implement an all-optical full-adder and full subtractor for triple input binary logic. If we combine the output of ports OP2, OP3, OP5 and OP8, we can obtain the "sum", whereas combination of output ports OP4, OP6, OP7 and OP8 gives the carry. Similarly, the combination of output ports OP2, OP3, OP5 and OP8 gives the "difference" whereas combination of output ports OP2, OP3, OP4 and OP8 also gives the "borrow" in case of full subtraction. Thus, this architecture works as an all-optical full adder and subtractor.

## 2. DESIGN OF DE-MULTIPLEXER AND MULTIPLEXER

A de-multiplexer is a circuit which is used to de-multiplex an incoming signal from one input channel to one of the many output channels via select lines. The select lines decide the routing of the incoming signal to the desired output channel. Fig.7 shows the architecture of a de-multiplexer based on BR coated microcavity switches S1, S2, and S3. Here X and Y act as select lines operating at wavelength 532 nm with a power of 200 µW. A laser source (LS) is coupled at port (1) of switch S1. Considering the detected output signal power ~ 200 μW, with an extinction of 9.8 dB [22], LS of 0.25 mW is enough to act as an incoming signal for the de-multiplexer. The cw light from the LS acts as an incoming signal to the switch S1. When the LS is switched OFF there is no light detected in any of the output ports of the architecture. When the LS is switched ON and both the select lines are zero i.e. X = 0 and Y = 0, no light is incident on the BR microcavity switches. BR is in its ground state hence, the incoming signal passes from the port (1) of switch S1 to port (1) of switch S3. As Y = 0, therefore light emerges from the port (2) of the switch S3 i.e. output port OP1 as shown in Table V.

Now, when X = 0 and Y = 1, light is incident on BR microcavity switches S2 and S3 respectively. It triggers the conformational changes in the BR. Thus, when an incoming signal is incident on port (1) of switch S1 it emerges from port (2) of switch S1, and reaches incoming port (1) of switch S3. Light at 532 nm is incident on switch S3; it redirects the incoming signal from port (1) to the port (3) of switch S3 via whispering gallery modes. The signal emerges from the output port OP2.

Considering the values taken by select lines as X = 1, Y = 0, when cw light is incident on the incoming port of switch S1, presence of the light (at 532 nm) at switch S1 triggers the switch to route the incoming signal from the port (1) to the port (3) of switch S1. The incoming signal reaches port (1) of the switch S2. As Y = 0, no light is incident on the switch and hence the incoming signal is directed to the output port OP3, via port (2) of switch S2. Thus light emerges from the output port OP3.

Similarly, when X = 1, and Y = 1 i.e. light (at 532 nm) is incident on all the switches. BR in all the microcavity switches will undergo conformational change and hence incoming signal is directed from port (1) to the port (3) of the switch S1. From there it reaches the incoming port (1) of switch S2. As Y = 1, the incoming signal is again rerouted to the port (3) of the switch S2 and hence to the output port OP4. Table V clearly depicts the various states of the routing signals of 1:4 de-multiplexer according to the select lines. To realize 1:4 de-multiplexer we need three switches. The scheme can be easily extended for n lines with $2^n-1$ switches. The whole circuit operates with very low power.

On the other hand multiplexer is a combinatorial circuit which is used to multiplex incoming signals from a number of input channels to a single output channel depending on the value of the select lines. Fig.8 shows a schematic of the architecture of a low power all-optical multiplexer. Consider IP1, IP2, IP3 and IP4 as incoming ports and X and Y as select lines. The incoming signals are multiplexed and sent to the output port according to the select lines. The output port acts as the port for the outgoing signal in the multiplexer case. The architecture can be expanded to multiplex n number of lines, for which we need $2^n-1$ switches. Table VI represents the various states of the multiplexer. It works in a similar manner as the de-multiplexer.

Let us consider the case when all the input ports have incoming signals i.e. IP1 = IP2 = IP3 = IP4 = LS, and X = 0, Y = 0. LS operate at wavelength 1311 nm. As Y = 0, incoming signal from incoming port IP1 moves to the port (2) of the switch S3, from there it is directed to the incoming port (1) of the switch S1. Here, X = 0 therefore, incoming signal from IP1 emerges from the port (2) of switch S1. Incoming signal is detected at the output port OP. Incoming signal from IP2 doesn't reach incoming port (1) of switch S1 as Y = 0. Incoming signal from IP3 moves from port (1) to port (2) of switch S2, but when it reaches switch S1, it is not directed to the output port as X= 0. Similarly, input signal from IP4 is not directed to the port (2) of the switch S2 as Y = 0 and it is not able to reach the output port OP of the switch S1.

Now, let us consider X = 0 and Y = 1 state. In this case, the incoming signal moves from port IP2 to the incoming port of the switch S1, as Y = 1, switch S3 directs the incoming signal from IP2 to the incoming port (1) of switch S1. Here X = 0, thus incoming signal flows from incoming port (1) to the port (2) of switch S1 and the signal is detected at output port OP. Incoming signal from IP1, IP3, and IP4 does not reach the output port of the switch S1.

When X = 1 and Y = 0, the incoming signal moves from port IP3 to the port (2) of the switch S2, from where it is directed to the output port OP of the switch S1 via whispering gallery modes. Other incoming signal from IP1, IP2, and IP4 does not reach output port of the switch S1. Similarly, when X = 1 and Y = 1, the incoming signal from IP4 is directed to the port (2) of the switch S2, from where it is directed to the output port OP of switch S1. Incoming signals from other ports are not directed to the output port OP of the switch S1. The select lines direct the flow of the desired incoming signal to the output channel and thus incoming signals can be easily multiplexed.

### 3. ARITHMETIC UNIT

Arithmetic unit (AU) is used to perform various arithmetic operations combined in one circuit. The number of functions performed by the AU depends on the number of select inputs. The operational principle of an all-optical arithmetic unit is explained by exploiting BR microcavity based all-optical switches, full-adders and multiplexers. The output of a BR based microcavity switch can be used as a control switch for the other. The optical circuit shown in the Fig. 9 can perform eight different arithmetic functions with two four bit numbers X ($X_3X_2X_1X_0$) and Y ($Y_3Y_2Y_1Y_0$). It gives the output Z ($Z_3Z_2Z_1Z_0$) depending on the value of three select inputs ($S_1$, $S_0$ and $C_{in}$). This scheme can be extended upto n bit numbers by using BR based microcavity switches, MUX and full-adders. Let us consider an example where X=1011 ($X_3X_2X_1X_0$) and Y = 1001 ($Y_3Y_2Y_1Y_0$). Eight different operations with four subsidiary functions are described and shown in Table VII in detail.

For AU to perform four bit operations as shown in the Fig.9 consist of BR based microcavity switches S1, S2,…, S13, multiplexers MUX-1,…, MUX-4, and full adders FA-1,…, FA-4 respectively. Select inputs $S_1$, $S_0$, $C_{in}$ operate at wavelength 532 nm. Input variable X also operates at wavelength 532 nm. Input variable Y operates with wavelength 1311 nm. Wavelength converters are used to realize the design of AU. Select inputs determine the arithmetic operation performed by the AU.

<u>Case: 1</u> a) When $S_1 = 0$, $S_0 = 0$ and $C_{in} = 0$, as the select lines are 0 hence input line IP0 of all the multiplexers (MUXs) will be active. IP0 of all the MUX is connected to the value of Y. It directs the value of $Y_0$ to the output port OP0. In the same way values of $Y_1$, $Y_2$ and $Y_3$ will be sent to the output ports OP1, OP2, OP3 respectively via. input ports.

The output of each MUX OP0, OP1, OP2, and OP3 is connected to one input (i.e. Y) of the full adder, FA-1, FA-2, FA-3 and FA-4 respectively. All the full adders receive the other input i.e. X directly. In this case $X_0$ is directly connected to one input (i.e. X) of the FA-1, carry in ($C_{in} = 0$). Here the select input $C_{in} = 0$ and the other two inputs $X_0 = 1$, $Y_0 = 1$. So $Z_0$ takes the value zero i.e. $Z_0 = 0$ and carry out i.e. $C_{out} = 1$. Now the carry out of the FA-1 is fed into the input of the FA-2. In this way $C_{in}$ of the full-adders FA-2, FA-3, FA-4, receive the value $C_{out}$ of the FA's FA-1, FA-2 and FA-3 respectively. So, in FA-2, $C_{in}=1$ and the other two inputs (X and Y) receive the value 1 and 0 (as $X_1 = 1$, $Y_1 = 0$). In this case output $Z_1 = 0$ and $C_{out} = 1$. Similarly in FA-3, $C_{in} = 1$, and the other two inputs (X and Y) receive the values 0 and 0 (as $X_2 = 0$ and $Y_2 = 0$). Hence $Z_2 = 1$ and $C_{out} = 0$. Finally in FA-4, $C_{in} = 0$, and the other two inputs (X and Y) receive the values 1 and 1 (as $X_3 = 1$ and $Y_3 = 1$), so $Z_3 = 1$ and $C_{out} = 1$. Here the output $Z_4$ takes the 1 as $C_{out} = 1$. The final output (Z) is 10100 ($Z_4 Z_3 Z_2 Z_1 Z_0$) which verifies the binary addition (X+Y) of two four bit numbers for select inputs $S_1 = 0$, $S_0 = 0$ and $C_{in} = 0$.

b) When $S_1 = 0$, $S_0 = 0$ and $C_{in} = 0$, and one of the input i.e. X = 0000 than the output is merely Y.

<u>Case: 2</u> a) When $S_1 = 0$, $S_0 = 0$ and $C_{in} = 1$. Here the select inputs are $S_1 = 0$, $S_0 = 0$, so all the operations are similar to that of case 1 a), with a difference that the carry in $C_{in}$ of the FA is 1 as the select input $C_{in} = 1$. The other two inputs receive the values 1 ($X_0 = 1$, $Y_0 = 1$). So, the output of FA-1 is 1 $Z_0 = 1$ and $C_{out} = 1$. Other FA's perform in a similar fashion and the final output is 10101 ($Z_4Z_3Z_2Z_1Z_0$). So it performs the operation of addition with carry.

b) When $S_1 = 0$, $S_0 = 0$ and $C_{in} = 1$, and one of the input i.e. $X = 0000$ than the function generated by the arithmetic unit is $Y+1$.

Case: 3 a) When $S_1 = 0$, $S_0 = 1$ and $C_{in} = 0$. As the select inputs $S_1 = 0$, $S_0 = 1$, the input data line IP1 will be selected for all the MUXs. Light form LS at wavelength 1311 nm is incident on switch S2. As the select input $Y_0 = 1$ the light emerges from the port P4. In this case output port P4 receives the light whereas P3 does not receive any light. So, P3 is in the zero state. i.e. P3 = 0. It is the complement of $Y_0$. Similarly, $Y_1$, $Y_2$, $Y_3$ acts as control input of the switches S5, S8 and S11. The ports P7, P11 and P15 receive the values that are the complement values of $Y_0$, $Y_1$, $Y_2$ and $Y_3$ respectively. So the complement values of $Y_0$, $Y_1$, $Y_2$ and $Y_3$ are fed into one input Y (i.e.Y) of full adder FA-1, FA-2, FA-3 and FA-4 respectively. Here $X_0$, $X_1$, $X_2$ and $X_3$ act as one input i.e. X of the FA's FA-1, FA-2, FA-3 and FA-4. Now in FA-1, carry in $C_{in} = 0$ and the input X receives the value 1 (as $X_0 = 1$), and the input Y receives the value 0, which is the complement of $Y_0$ (as $Y_0 = 1$). According to the operation principle of the full adder the output sum of FA-1 is one (i.e. $Z_0 = 1$) and the carry out is zero (i.e. $C_{out} = 0$). Again in FA-2, $C_{in} = 0$ (as $C_{out}$ of the FA-1 is 0). The input X receives the value 1 (as $X_1 = 1$) and the input Y receives the value 1, which is the complement of $Y_1$ (as $Y_1 = 0$). So, $Z_1 = 0$ and $C_{out} = 1$. Similarly in FA-3, $C_{in} = 0$ and the input X receives the value 0 (as $X_2 = 0$), and the input Y receives the value 1 which is the complement of $Y_2$ (as $Y_2 = 0$). Hence IP2 = 0, and $C_{out}$ =1. Finally in FA-4, $C_{in} = 1$ i.e. $Z_4 = 1$. So the final output Z is 10001 ($Z_4Z_3Z_2Z_1Z_0$) which also verifies the third operation ($X + Y'$) as given in table for select inputs $S_1 = 0$, $S_0 = 1$ and $C_{in} = 0$.

b) When $S_1 = 0$, $S_0 = 1$ and $C_{in} = 0$, and one of the input i.e. $X = 0000$, the function generated at the output is $Y'$.

<u>Case: 4</u> a) When $S_1 = 0$, $S_0 = 1$ and $C_{in} = 1$. All the operations are similar to that of case 3 a), with a difference that the carry in of the full adder FA-1 is 1 as the select input $C_{in} = 1$. Therefore, the input X receives the value 1 (as $X_0 = 1$) and the input Y receive the value 0, which is the complement of $Y_0$ (as $Y_0 = 1$). So output of $Z_0$ of FA-1 is zero i.e. $Z_0 = 0$, and the carry out is one i.e. $C_{out} = 1$. Other output Z as 10100. In this case the optical circuit uses 2's complement method of subtraction. As the final carry ($C_{out}$) from FA-4 is one (i.e. $Z_4 = 1$) that means the result will be positive and hence the final carry is to be discarded. Hence, the final result is positive as its value is 0010 ($Z_3Z_2Z_1Z_0$), which also verifies the 2's complement subtraction operation (X + Y' + 1) for select inputs ($S_1 = 0$, $S_0 = 1$ and $C_{in} = 1$) as given in table. Now if we consider X = 1001 ($X_3X_2X_1X_0$) and Y = 1011 ($Y_3Y_2Y_1Y_0$), then the optical circuit generates the output Z as 01110 ($Z_4Z_3Z_2Z_1Z_0$). As the final carry is 0 (i.e. $Z_4 = 0$) that means the result will be negative and hence the final carry $Z_4$ is to be discarded. Also taking the 2's complement of 1110 ($Z_3Z_2Z_1Z_0$). So, the final result is negative as its value is 1110 (2's complement of 1110).

b) When $S_1 = 0$, $S_0 = 1$ and $C_{in} = 0$, and one of the input i.e. X = 0000, the function generated at the output is Y' + 1.

<u>Case: 5</u> When $S_1 = 1$, $S_0 = 0$ and $C_{in} = 0$. As the select inputs $S_1 = 1$, $S_0 = 0$, the input data line IP2 will be selected for all the MUXs. Light from LS is incident on switch S1. According to the switching principle, Port P1 will receive the value, which will be the complement of $C_{in}$ (i.e. $C_{in}'$). Again $X_0$, $X_1$, $X_2$ and $X_3$ act as control inputs for switches S3, S6, S9 and S12 respectively. Port P1 (i.e. $C_{in}'$) is connected to switches S4, S7, S10 and S13. According to the switching principle, ports P6, P10, P14 and P18 receive the values which will be equal to $XC_{in}'$ and connected to IP2 line of all the MUXs receiving the values equal to $X_0$, $X_1$, $X_2$ and $X_3$ respectively (because $XC_{in}' = X$ as $C_{in} = 0$), and also generate the values as $X_0$, $X_1$, $X_2$ and $X_3$ respectively. Now in FA-1 carry in i.e. $C_{in} = 0$ (as select inputs $C_{in} = 0$), and the other two inputs (X and Y) receive the same value (as $X_0 = 1$ as independent of $Y_0$). So, the FA-1 generates the output as zero i.e. $Z_0 = 0$, and the carry out $C_{out}$ as one i.e. $C_{out} = 1$. Again in FA-2 $C_{in} = 1$ and the other two inputs receive the same value 1 (as $X_1 = 1$). So in this case $Z_1$ takes the value 1 i.e. $Z_1 = 1$ and the carry out receives the value 1 i.e. $C_{out} = 1$. Similarly in FA-3 and FA-4 values of X and Y propagate in the same manner. So the final output Z is 10110 ($Z_4Z_3Z_2Z_1Z_0$), which verifies the double operation (2X) of a four bit number (X) for select inputs ($S_1 = 1$, $S_0 = 0$ and $C_{in} = 0$) as given in Table VII.

<u>Case: 6</u> When $S_1 = 1$, $S_0 = 0$ and $C_{in} = 1$. Here also the select inputs $S_1 = 1$, $S_0 = 0$, the input data line IP2 will be selected for all the MUXs. As the select inputs $C_{in} = 1$, the input data line IP2 of all the MUXs receive the value 0. ($XC_{in} = 0$ as select input $C_{in} = 1$). Hence one input i.e. Y of all the full adders receives the value 0. So, in FA-1, $C_{in} = 1$ (as the select input $C_{in} = 1$), and the two other inputs (X and Y) receive the values 1 and 0 (as $X_0 = 1$ and $X_0C_{in}' = 0$. So the output of the FA-1 is 0 i.e. $Z_0 = 0$ and the carry out $C_{out}$ is one. Again in FA-2, $C_{in} = 1$, and the other two inputs (X and Y) receive the values 1 and 0 (as $X_1 = 1$ and $X_1C_{in}' = 0$). So in this case $Z_1$ takes the value 0, $Z_1 = 0$ and the carry out receives the value 1 i.e. $C_{out} = 1$. Similarly in FA-3, $C_{in} = 1$, and the other two inputs (X and Y) receives the values 0 and 0 (as $X_2 = 0$ and $X_2C_{in}' = 0$). Hence $Z_2 = 1$ and $C_{out} = 0$. Finally in FA-4, $C_{in} = 0$, and the other two inputs (X and Y) receive the values 1 and 0 i.e. $Z_4 = 0$. So, the final output Z is 01100 ($Z_4Z_3Z_2Z_1Z_0$), which also verifies the increment operation (X + 1) of a four bit number (X) for select inputs ($S_1 = 1$, $S_0 = 0$ and $C_{in} = 1$) as given in Table VII.

<u>Case: 7</u> When $S_1 = 1$, $S_0 = 1$ and $C_{in} = 0$. As the select inputs $S_1 = 1$, $S_0 = 1$, the input data line IP3 will be selected for all the MUXs. Here the IP3 line of all the MUXs is connected to the control signal which is always in one (high) state. So all the MUXs generate the output as 1. Thus one input (Y) of all the full adders receives the value 1. Now in FA-1, $C_{in} = 0$ (as the select input $C_{in} = 0$), and input X receives the value 1 ($X_0 = 1$) and input Y receives the value 1 (as MUX-1 generates output 1). According to the operational principle of the full adder, it generates the output as zero i.e. $Z_0 = 0$ and the carry out as one i.e. $C_{out} = 1$. Again in FA-2, $C_{in} = 1$ and input X receives the value 1 (as MUX-2 gives output 1). In this case it gives output as zero, i.e. $Z_1 = 1$ and the carry out i.e. $C_{out} = 1$. Similarly in FA-3, $C_{in} = 1$, and input X receives the value 0 (as $X_2 = 0$) and input Y receives the value 1 (as MUX-3 generates output 1). Hence the output $Z_2 = 0$ and $C_{out} = 1$. Finally in FA-4 $C_{in} = 1$, and input X receives the value 1 (as $X_3 = 1$) and input Y receives the value 1 (as MUX-4 gives output 1). So $Z_3 = 1$ and $C_{out} = 1$, i.e. $Z_4 = 1$. The circuit generates the output Z as 11010 ($Z_4Z_3Z_2Z_1Z_0$). In this case circuit uses 2's complement method for subtraction for decrement (X - 1) of a four bit number for select inputs ($S_1 = 1$, $S_0 = 1$, $C_{in} = 0$) as given in the Table VII.

<u>Case: 8</u> When $S_1 = 1$, $S_0 = 1$ and $C_{in} = 1$. Here also the select inputs are 1: so all the operations are similar to that of case 7 with a difference that the carry in $C_{in}$ of the FA-1, $C_{in} = 1$ (as the select input $C_{in} = 1$), and input X receives the value 1 (as $X_0 = 1$) and input Y receives the value 1 (as MUX-1 generates output 1), it generates the output one, i.e. $X_0 = 1$ and the carry out is one i.e. $C_{out} = 1$. Again in FA-2, $C_{in} = 1$ and the other two inputs (X and Y) receive the values 1 and 1 (as $X_1 = 1$ and MUX-2 gives output = 1). In this case it gives the output 1, i.e. $Z_0 = 1$ and carry out i.e. $C_{out} = 1$. Similarly in FA-3 and FA-4 the value of X and Y propagate in the same manner. The circuit gives the output Z as 11011 ($Z_4Z_3Z_2Z_1Z_0$). As Z ($Z_4Z_3Z_2Z_1Z_0$) is a four bit number and the final carry is to be discarded. So the final result is 1101 ($Z_3Z_2Z_1Z_0$) which verifies the transfer operation of a four bit number for select inputs ($S_1 = 1$, $S_0 = 1$ and $C_{in} = 1$) as given in Table VII. In this way we can perform twelve (eight main and four subsidiary) operations.

## IV. RESULTS AND DISCUSSION

All-optical switching in BR coated microsphere has been demonstrated experimentally. In the BR coated microcavity switch, the switching of incoming signal operating at wavelength 1310 nm between the output ports (2 and 3) is photo-induced with a fiber-coupled green pump laser (at 532 nm) which controls the conformational state of the adsorbed BR. In order to achieve fast switching, blue light beam is injected into the BR coated microcavity switch in synchronization with a green pump laser beam (at 532 nm). The blue light beam operating at wavelength 405 nm helps in truncating the photocycle of the BR molecules at the M intermediate state, which is near to its peak absorption wavelength of 410 nm.

This basic all-optical BR coated microcavity switch has been used to design higher computing circuits. All the designs presented in this paper are all-optical in nature. The proposed designs are simple and can easily implement multiple functions. The designs have good scalability. The proposed design of AU does not require erbium-doped fibre amplifiers for signal regeneration and uses less number of wavelength converters. The switching of near IR wavelength (1310 nm) probe with the controlled pump operating at 532 nm may find potential applications in the telecommunication as well. The microresonator cavity can be tailored to effectively switch at a desired wavelength. Moreover, with the powerful capabilities of nano-biotechnological techniques, the desired response of BR molecules can also be tailored to meet device specifications. Cascading can be easily done as BR based microcavity switches dissipate very less power. Moreover, Q-factor of these microcavity switches is very high which makes these designs highly sensitive. Coating the microcavities with a photosensitive material is of critical value, as the material should exhibit high sensitivity, absorption, fast dynamics, photo and thermal stability and potential to tailor its properties. BR protein is a natural photochromic material that exhibits this unique combination of properties for practical realization. Although proposed designs can be realized with dual-ring switches and other interferometric configurations, single BR based microcavity switch has advantages in terms of simple geometry, ease of fabrication, high thermal and photo stability, high fan out, cost-effectiveness and low power operation.

## V. CONCLUSION

In this paper, we have shown experimentally all-optical switching in BR coated microcavity switch. Blue light beam (at 405 nm) is injected into the BR coated microcavity switch in synchronization with a green pump laser beam (at 532 nm) for fast switching. The blue light beam operating at wavelength 405 nm helps in truncating the photocycle of the BR molecules at the M intermediate state, which is near to its peak absorption wavelength of 410 nm. We have also presented a theoretical scheme which combines the advantages of both BR based microcavity switch and tree architecture for designing low power all-optical, half adder/subtractor, de-multiplexer, multiplexer and arithmetic unit. These schemes can easily be extended and implemented for any higher number of input digits, by proper interconnection of BR based microcavity switches using vertical and horizontal extension of the tree and by suitable branch selection. Due to the high-Q factor and small dimensions, switching at low power is possible with these BR coated microcavities. They represent an alternative to the waveguide based techniques providing exceptional sensitivity and straightforward optical integration on micron scales. The proposed designs can yield large computing circuits and networks within mW-W power budget. The architectures with BR microcavity switches have an advantage that they require less number of switches and have low power consumption as compared to other tree based architectures for realization of these higher computing circuits. To realize n bit computation the proposed design requires only $2^n-1$ switches.

The change in the refractive index in a BR based microcavity switch depends on the wavelength of the control beam making the switches tunable. The combined advantages of high Q-factor, tunability, compactness and low power control signals of BR coated microcavities along with the flexibility of cascading switches in tree architectures to form circuits makes the designs promising for practical applications. The proposed designs proposes a new paradigm for all-optical computing based on hybrid nano-bio-photonic photonic integrated devices employing biomolecules to perform photonic functions.

TABLE I. TRUTH TABLE OF FIG. 2

| Incoming Signal | Control Signal | Port (3) | Port (2) |
|---|---|---|---|
| 0 | 0 | 0 | 0 |
| 0 | 1 | 0 | 0 |
| 1 | 0 | 0 | 1 |
| 1 | 1 | 1 | 0 |

TABLE II. STATE OF DIFFERENT OUTPUT PORTS FOR DIFFERENT VALUES OF X

AND Y IN A TREE ARCHITECTURE.

| Input | | Output at different Ports | | | |
|---|---|---|---|---|---|
| X | Y | OP1 (X′Y′) | OP2 (X′Y) | OP3 (XY′) | OP4 (XY) |
| 0 | 0 | 1 | 0 | 0 | 0 |
| 0 | 1 | 0 | 1 | 0 | 0 |
| 1 | 0 | 0 | 0 | 1 | 0 |
| 1 | 1 | 0 | 0 | 0 | 1 |

TABLE III. TRUTH TABLE OF HALF-ADDER AND SUBTRACTOR.

| Input | | Output of Half adder | | Output of Half Subtractor | |
|---|---|---|---|---|---|
| X | Y | Sum | Carry | Difference | Borrow |
| 0 | 0 | 0 | 0 | 0 | 0 |
| 0 | 1 | 1 | 0 | 1 | 1 |
| 1 | 0 | 1 | 0 | 1 | 0 |
| 1 | 1 | 0 | 1 | 0 | 0 |

TABLE IV. TRUTH TABLE OF FULL ADDER AND SUBTRACTOR.

| Input | | | Output | | | |
|---|---|---|---|---|---|---|
| X | Y | Z | Full-Adder | | Full-Subtractor | |
| | | | Sum | Carry | Difference | Borrow |
| 0 | 0 | 0 | 0 | 0 | 0 | 0 |
| 0 | 0 | 1 | 1 | 0 | 1 | 1 |
| 0 | 1 | 0 | 1 | 0 | 1 | 1 |
| 0 | 1 | 1 | 0 | 1 | 0 | 1 |
| 1 | 0 | 0 | 1 | 0 | 1 | 0 |
| 1 | 0 | 1 | 0 | 1 | 0 | 0 |
| 1 | 1 | 0 | 0 | 1 | 0 | 0 |
| 1 | 1 | 1 | 1 | 1 | 1 | 1 |

TABLE V. TRUTH TABLE OF 1: 4 DE-MULTIPLEXER

| Incoming Signal | Select Inputs/Control Signals | | Data Outputs | | | |
|---|---|---|---|---|---|---|
| LS | X | Y | OP1 | OP2 | OP3 | OP4 |
| 0 | 0 | 0 | 0 | X | X | X |
| 1 | 0 | 0 | 1 | X | X | X |
| 0 | 0 | 1 | X | 0 | X | X |
| 1 | 0 | 1 | X | 1 | X | X |
| 0 | 1 | 0 | X | X | 0 | X |
| 1 | 1 | 0 | X | X | 1 | X |
| 0 | 1 | 1 | X | X | X | 0 |
| 1 | 1 | 1 | X | X | X | 1 |

TABLE VI. TRUTH TABLE FOR 4 : 1 MULTIPLEXER

| Data Inputs | | | | Select Inputs/Control Signals | | Outgoing Signal |
|---|---|---|---|---|---|---|
| IP1 | IP2 | IP3 | IP4 | X | Y | O/S |
| 0 | X | X | X | 0 | 0 | 0 |
| 1 | X | X | X | 0 | 0 | 1 |
| X | 0 | X | X | 0 | 1 | 0 |
| X | 1 | X | X | 0 | 1 | 1 |
| X | X | 0 | X | 1 | 0 | 0 |
| X | X | 1 | X | 1 | 0 | 1 |
| X | X | X | 0 | 1 | 1 | 0 |
| X | X | X | 1 | 1 | 1 | 1 |

TABLE VII. ARITHMETIC FUNCTIONS REALIZED USING ALL-OPTICAL AU

| Functions Generated | Select Inputs | | | Inputs : X, Y |
|---|---|---|---|---|
| | S1 | S0 | $C_{in}$ | Output : Z ($Z_4 Z_3 Z_2 Z_1 Z_0$) |
| Binary-addition | 0 | 0 | 0 | Z = X + Y, <br> Z = Y; if X = 0 |
| Addition with Carry | 0 | 0 | 1 | Z = X + Y + 1, <br> Z = Y + 1; if X = 0 |
| Subtract with borrow | 0 | 1 | 0 | Z = X + Y', <br> Z = Y'; if X = 0 |
| Subtract (2's complement) | 0 | 1 | 1 | Z = X + Y' + 1, <br> Z = Y' + 1; if X = 0 |
| Double of X | 1 | 0 | 0 | Z = 2 X |
| Increment of X | 1 | 0 | 1 | Z = X + 1 |
| Decrement of X | 1 | 1 | 0 | Z = X – 1 |
| Transfer of X | 1 | 1 | 1 | Z = X |

**FIGURE CAPTIONS**

1.  Schematic of the photochemical cycle of BR molecule. Subscripts indicate absorption peaks in nm. Solid and dashed arrows represent thermal and photo-induced transitions respectively.

2.  Schematic of BR coated microcavity switch.

3.  Experimental setup of all-optical switching in BR coated microcavity switch.

4.  Variation of the signal probe beam (operating at 1310 nm) with the modulating green and blue laser pump beams operating at wavelengths 532 nm and 405 nm respectively.

5.  Design of an all-optical half-adder/half-subtractor circuit (B.S = Beam Splitter, B.C = Beam Combiner, LS = Laser Source)

6.  Design of an integrated all-optical full-adder and subtractor circuit (B.S = Beam Splitter, *B.C = Beam Combiner, OP = Output Port, LS = Laser Source).

7.  Design of de-multiplexer with BR based microcavity switch (LS = laser source).

8.  Design of multiplexer with BR based microcavity switch (B.S = Beam Splitter, LS = Laser Source).

9.  Design of all-optical arithmetic unit (MUX = Multiplexer, FA = Full-Adder).

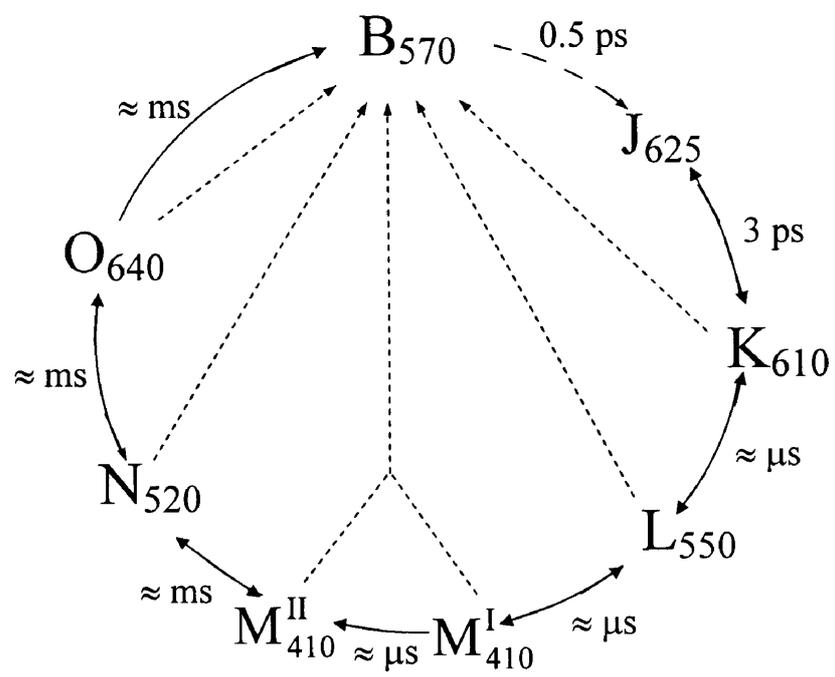

Fig.1.

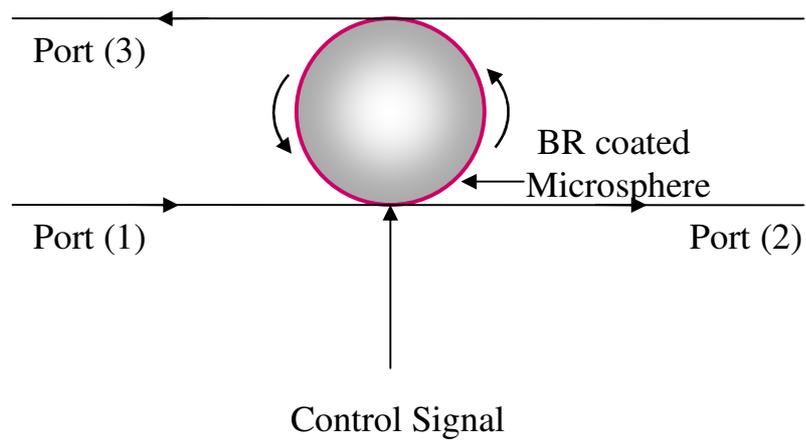

Port (3)

Port (1)                                                    Port (2)

BR coated
Microsphere

Control Signal

Fig.2

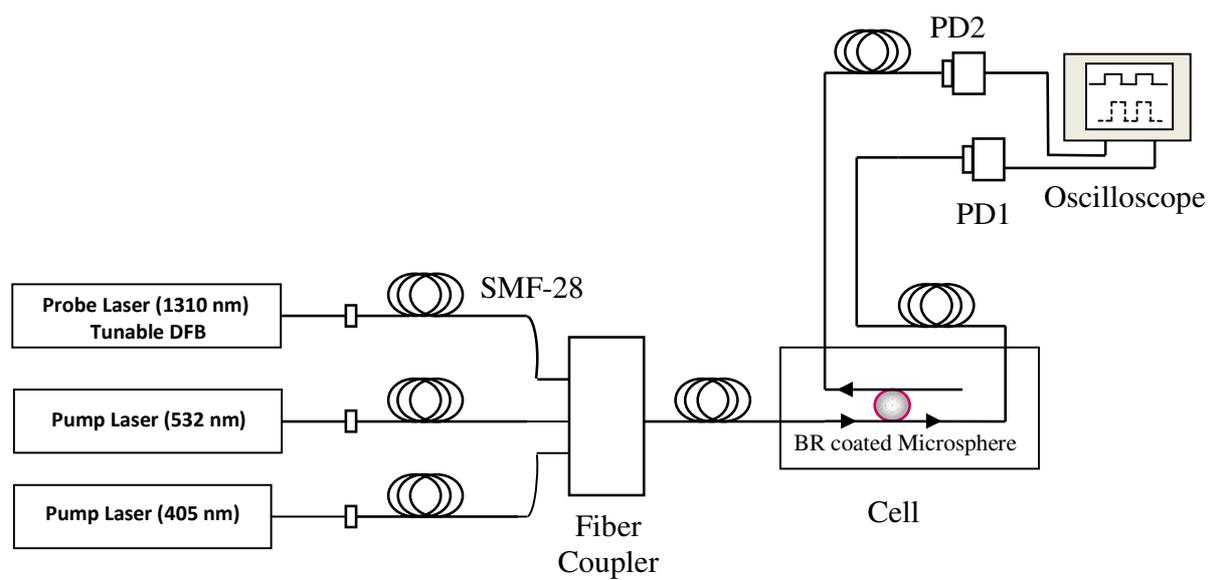

Fig. 3

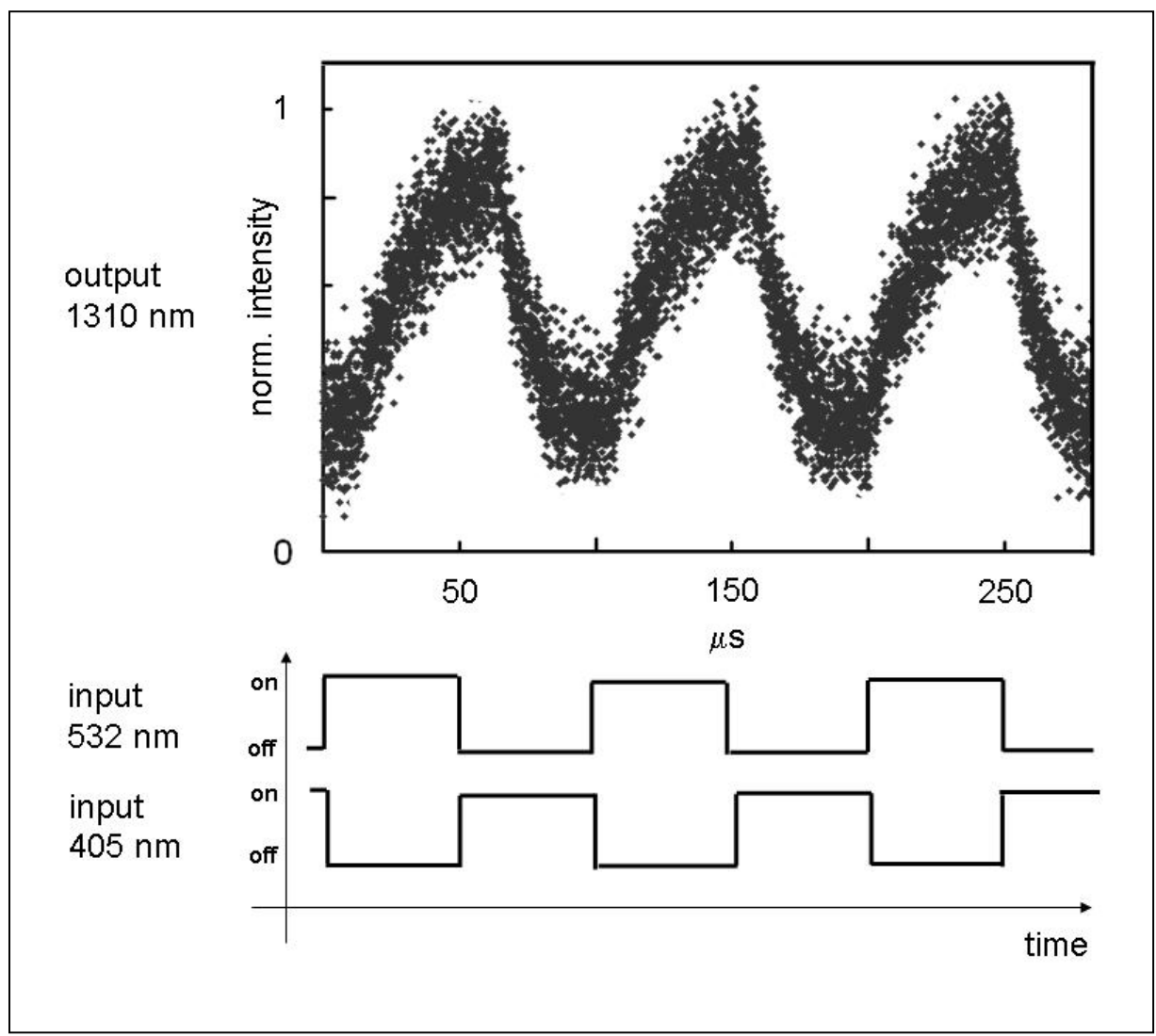

Fig. 4

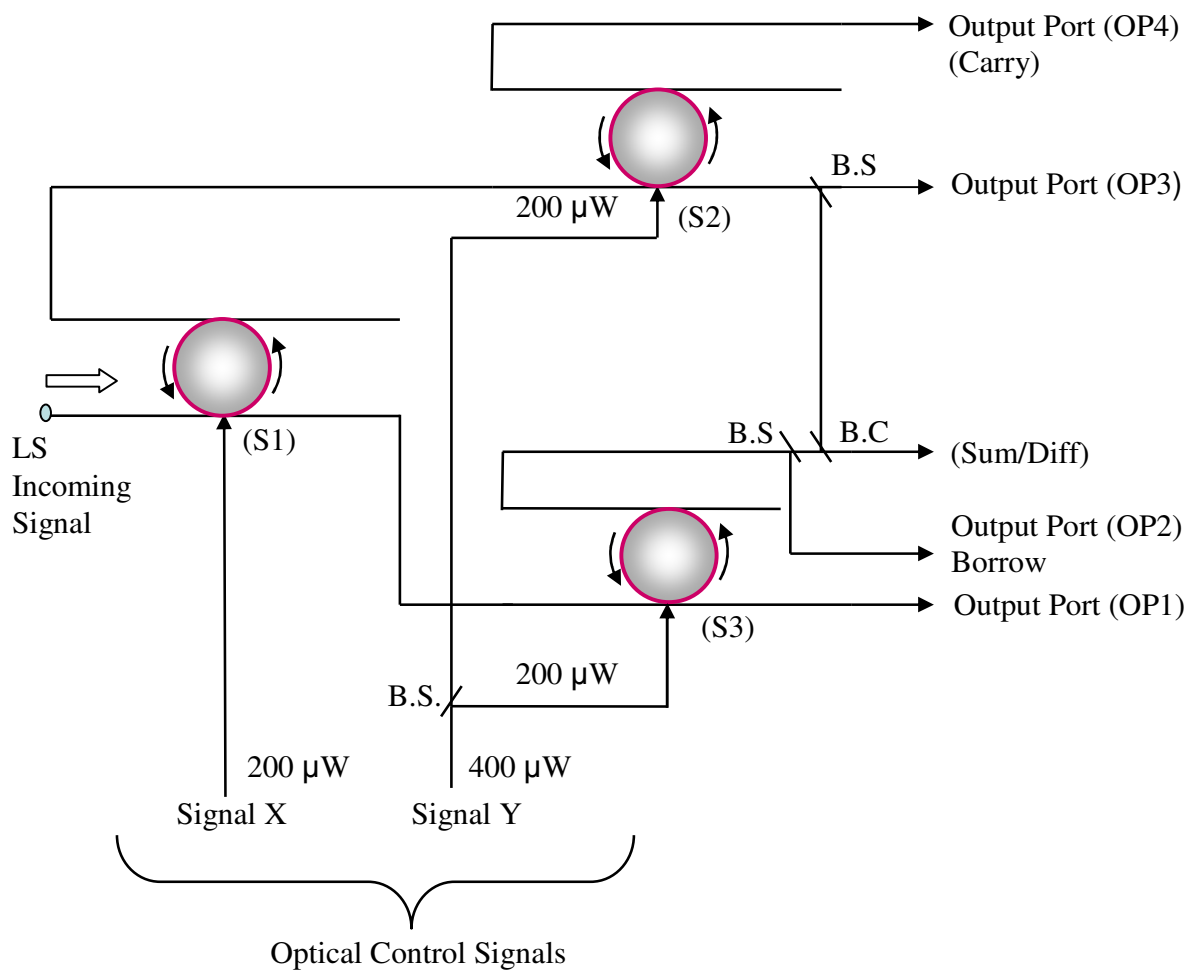

Fig. 5

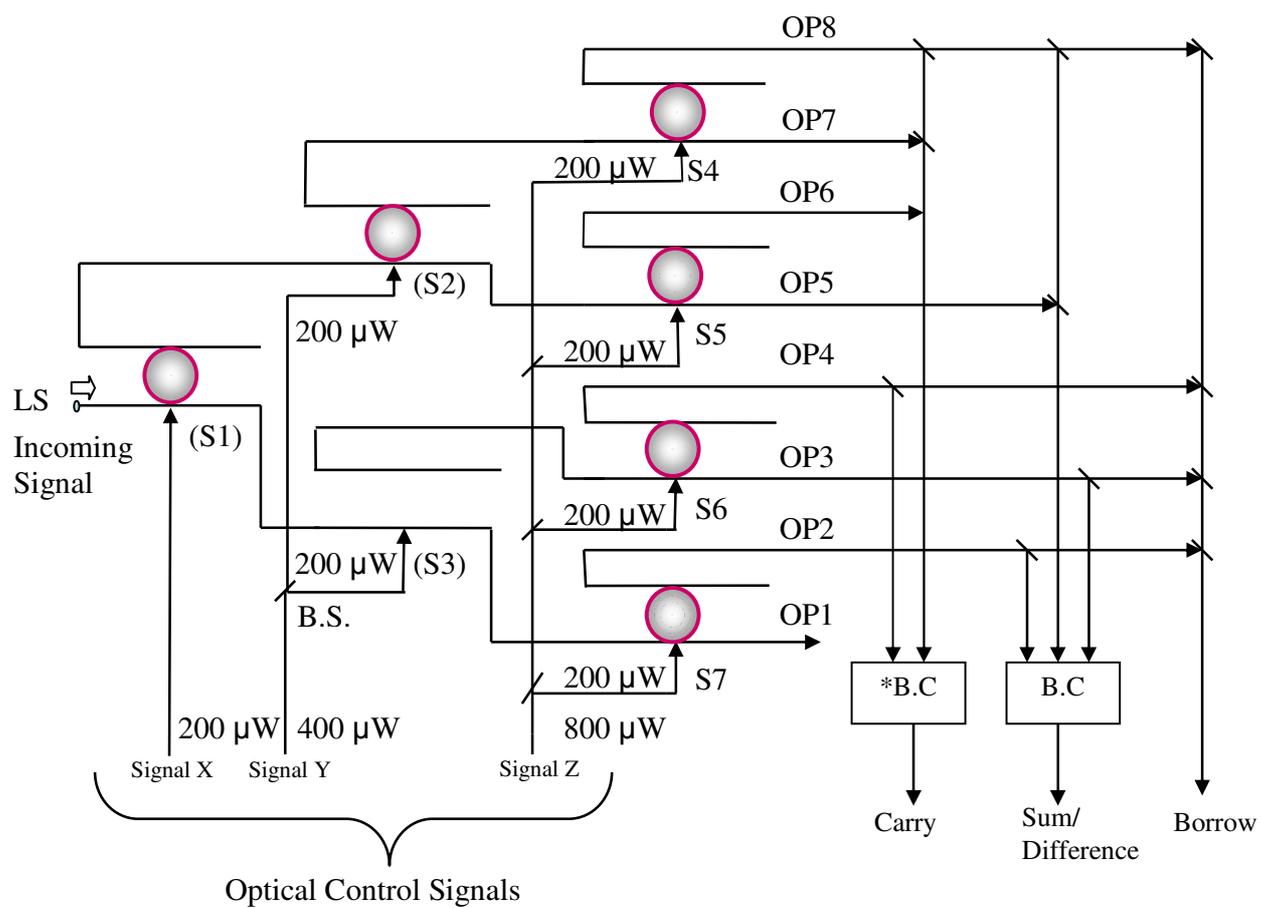

Optical Control Signals

*B.C = Beam Combiner, OP = Output Port        Fig. 6

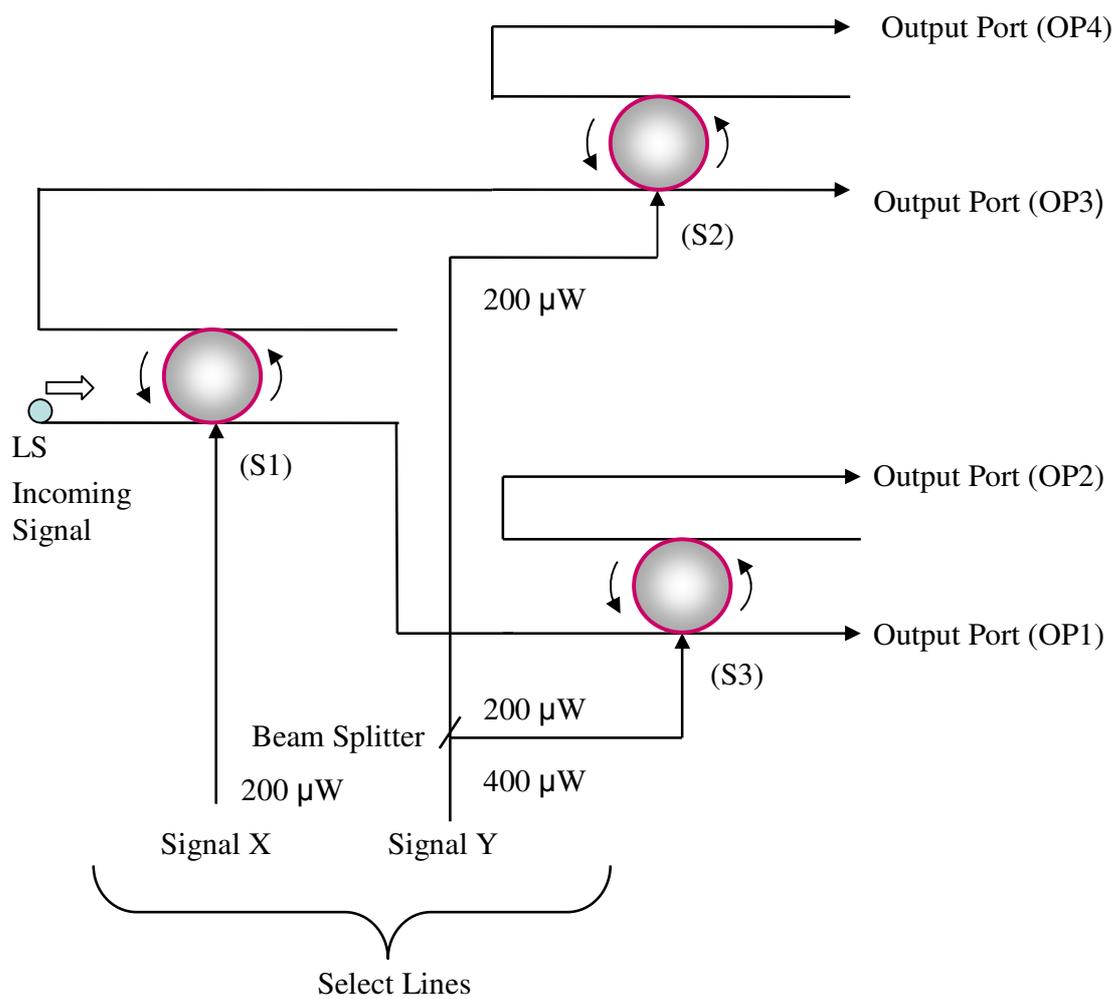

Output Port (OP4)

Output Port (OP3)

(S2)

200 μW

Output Port (OP2)

Output Port (OP1)

(S3)

200 μW

LS

Incoming
Signal

(S1)

Beam Splitter

200 μW

400 μW

Signal X     Signal Y

Select Lines

Fig. 7

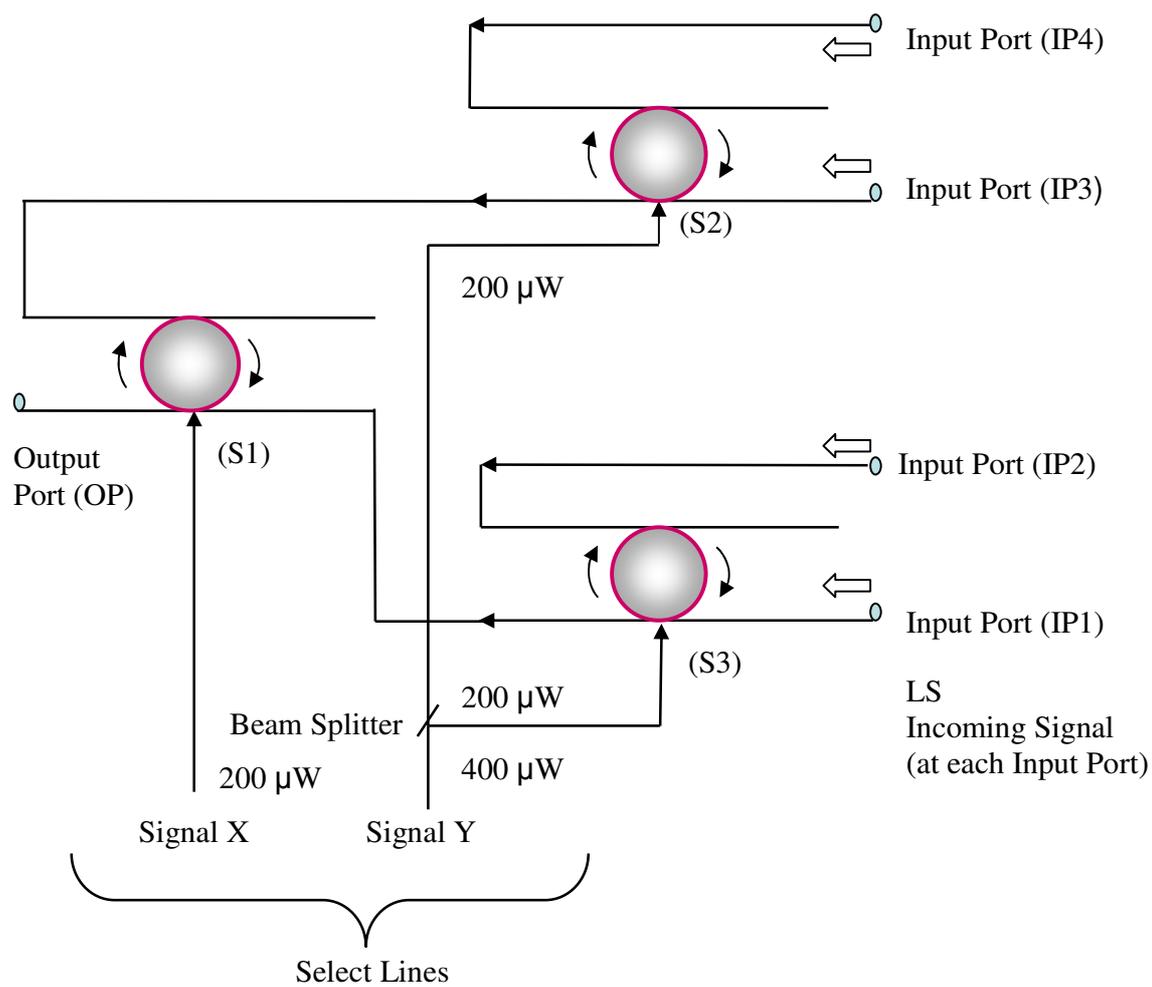

Input Port (IP4)

Input Port (IP3)

(S2)

200 µW

Output
Port (OP)

(S1)

Input Port (IP2)

Input Port (IP1)

(S3)

LS
Incoming Signal
(at each Input Port)

Beam Splitter

200 µW

200 µW

400 µW

Signal X

Signal Y

Select Lines

Fig. 8

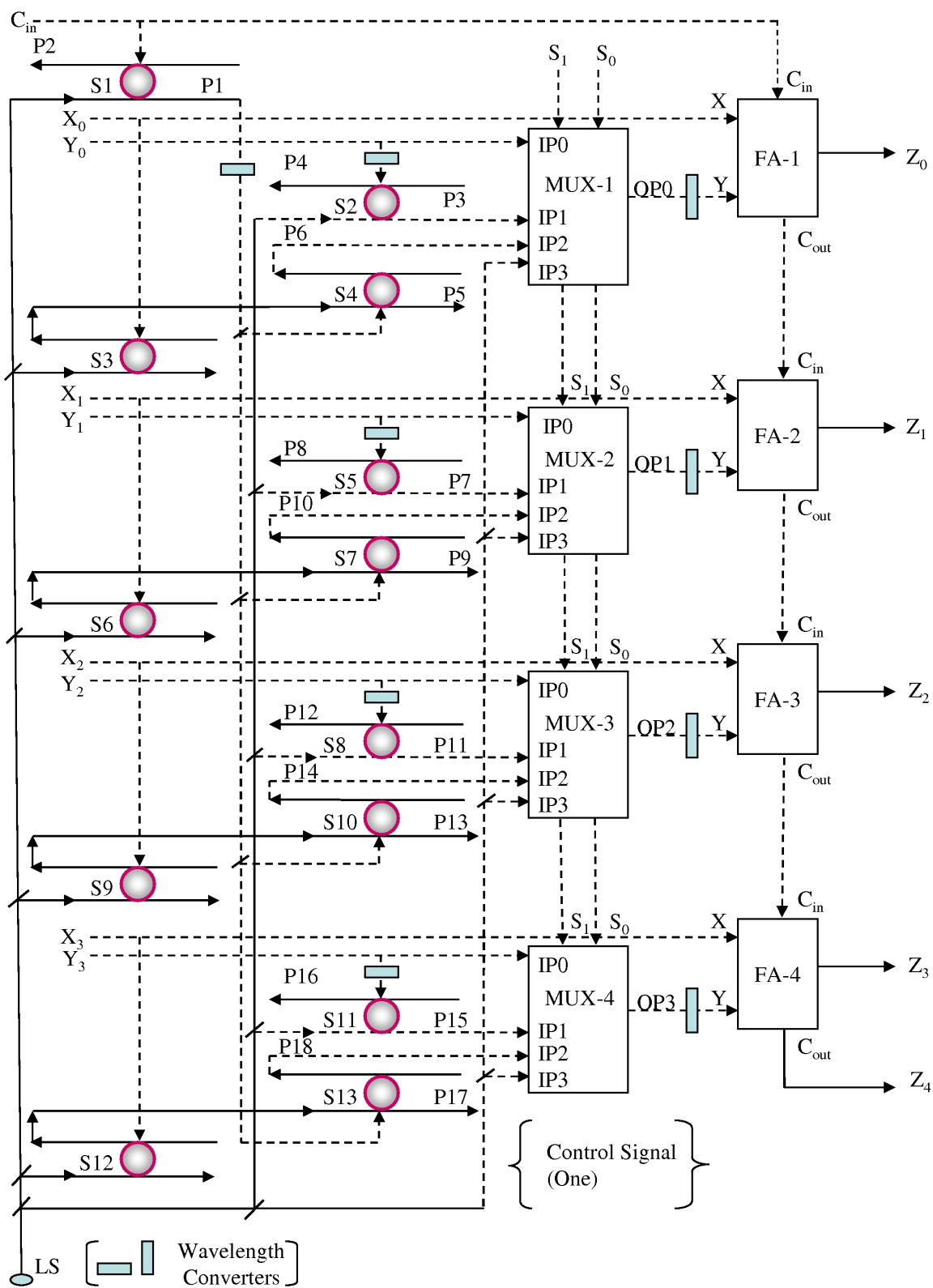

Fig. 9